\documentclass[prb,twocolumn]{revtex4}
\usepackage{graphicx}

\begin{document}

\title{Effect of pressure on the polarized infrared optical response of quasi-one-dimensional LaTiO$_{3.41}$}

\author{S. Frank,$^{1}$ C. A. Kuntscher,$^{1*}$ I. Loa,$^{2}$ K. Syassen,$^{2}$ and F. Lichtenberg$^3$}
\address{
$^1$ 1. Physikalisches Institut, Universit\"at Stuttgart, Pfaffenwaldring 57, D-70550 Stuttgart, Germany\\
$^2$ Max-Planck-Institut f\"ur Festk\"orperforschung, Heisenbergstrasse\ 1, D-70569 Stuttgart, Germany \\
$^3$ Experimentalphysik VI, Institut f\"ur Physik, EKM,
Universit\"at Augsburg, Universit\"atsstrasse 1, D-86135
Augsburg, Germany}

\date{\today}

\begin{abstract}
The pressure-induced changes in the optical properties of the
quasi-one-dimensional conductor LaTiO$_{3.41}$ were studied by
polarization-dependent mid-infrared micro-spectroscopy at room
temperature. For the polarization of the incident radiation
parallel to the conducting direction, the optical conductivity
spectrum shows a pronounced mid-infrared absorption band,
exhibiting a shift to lower frequencies and an increase in
oscillator strength with increasing pressure. On the basis of
its pressure dependence, interpretations of the band in terms
of electronic transitions and polaronic excitations are
discussed. Discontinuous changes in the optical response near
15~GPa are in agreement with a recently reported
pressure-induced structural phase transition and indicate the
onset of a dimensional crossover in this highly anisotropic
system.
\end{abstract}


\maketitle

\section{Introduction}
Perovskite-related titanates are intensively studied both
experimentally and theoretically, since they show interesting
physical properties due to the complex interplay of structural and
electronic degrees of freedom. Recently, special attention was
paid to the compounds $R$TiO$_{3.0}$ (with $R$ being a trivalent
rare-earth ion), \cite{Mochizuki04a} which are Mott-Hubbard
insulators with a Ti 3d$^1$ configuration. LaTiO$_{3.0}$, in
particular, has attracted much attention
[\onlinecite{Mochizuki04a,Khaliullin00,Keimer00,Cwik03,Mochizuki03,Hemberger03,Kiyama03,Pavarini04,Haverkort05,Ruckkamp05,Ulrich05}],
partly because of different orbital ordering scenarios.
LaTiO$_{3.0}$ is the end member of the series LaTiO$_{3.5-x}$ with
0$\leq$$x$$\leq$0.5, in which one finds a rich variety of
different structural, magnetic, and electronic properties,
depending on the composition parameter $x$. Besides the
Mott-Hubbard system LaTiO$_{3.0}$, the band insulator
LaTiO$_{3.5}$ is a prominent member of the series because of its
ferroelectricity up to extremely high
temperatures.\cite{Nanamatsu74}

\begin{figure}[t]
\includegraphics[width=0.9\columnwidth]{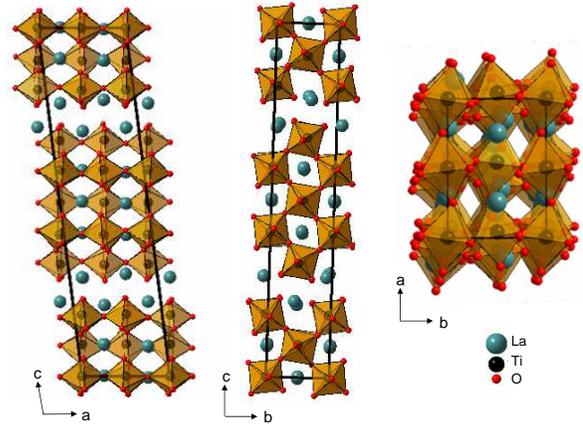}
\caption{(Color online) Crystal structure of LaTiO$_{3.41}$
composed of perovskite-like slabs of vertex-sharing TiO$_6$
octahedra, which are separated by additional oxygen
layers.\cite{Daniels03} Along the $a$ axis the TiO$_6$ octahedra
are connected via their apical oxygen atoms forming chains.}
\label{structure}
\end{figure}

Other compounds of the series LaTiO$_{3.5-x}$ were synthesized
recently,\cite{Lichtenberg01} among them LaTiO$_{3.41}$, which is
a quasi-one-dimensional (quasi-1D) conductor according to its
anisotropic DC resistivity and infrared
response.\cite{Lichtenberg01,Kuntscher03} It crystalizes in a
monoclinic structure (space group P2$_1$/c) with lattice
parameters $a$=7.86 \AA, $b$=5.53 \AA, $c$=31.48 \AA, and
$\beta$=97.1$^{\circ}$. \cite{Daniels03}  The structure consists
of slabs of vertex-sharing TiO$_6$ octahedra separated by
additional oxygen layers (see Fig.\ \ref{structure}). Along $c$
the slabs are five octahedra wide, and neighboring slabs are
shifted along the $a$ axis by half an octahedron. The octahedra
are tilted away from the $a$ axis and rotated around this axis,
similarly to the GdFeO$_3$-type arrangement of octahedra in
LaTiO$_{3.0}$.\cite{Cwik03} LaTiO$_{3.41}$ can thus be viewed as
being built of LaTiO$_{3.0}$-type slabs.
The characteristic units of the crystal structure are chains of
TiO$_6$ octahedra, connected via their apical oxygen atoms and
oriented along the $a$ axis. These chains can serve as an
explanation for the anisotropic electronic transport
properties of LaTiO$_{3.41}$, with relatively low resistivity
values along the chain direction $a$.\cite{Lichtenberg01}

To shed light on the conduction mechanism of quasi-1D conducting
LaTiO$_{3.41}$ its polarization-dependent reflectivity was studied
as a function of temperature.\cite{Kuntscher03} An
anisotropic optical response was observed, with a Drude
contribution of free carriers for the polarization of the incident
radiation parallel to the conducting crystal axis $a$
and an insulating character for the perpendicular direction $b$.
Furthermore, the {\bf E}$\parallel$$a$ optical conductivity spectrum
includes a pronounced mid-infrared (MIR) absorption band, showing a
shift to lower frequencies and an increase in oscillator strength
with decreasing temperature. A polaronic model was proposed to account
for the temperature dependence of the MIR band.\cite{Kuntscher03}

The application of external pressure to LaTiO$_{3.41}$ up to
$P$=18~GPa leads to continuous changes of the crystal
structure:\cite{Loa04} The axial compressibilities are
anisotropic with a ratio of approximately 1:2:3 for the $a$,
$b$, and $c$ axes. The large compressibility along $c$ results
from the highly compressible oxygen-rich layers separating the
LaTiO$_{3.0}$-type slabs (see Fig.\ \ref{structure}). The
differences in axis compressibilities cause a small increase of
the monoclinic angle from 97.17$^\circ$ to 97.43$^\circ$ with
increasing pressure up to 18~GPa. From the pressure dependence
of the lattice parameters the octahedral tilt angle against the
$a$ axis was estimated to double at 18~GPa compared to ambient
conditions. Above 18~GPa the appearance of additional
reflections in the x-ray diffraction diagrams indicate a
sluggish structural phase transition, which is completed at 24
GPa.\cite{Loa04}

In this paper we report the effect of pressure on the MIR
reflectivity of LaTiO$_{3.41}$. Reflectivity spectra of a
single-crystal sample were measured for polarizations along the
$a$ and $b$ axes using MIR micro-spectroscopy in combination
with a diamond anvil high pressure cell. The primary motivation
is twofold: (1) Pressure effects are of considerable interest
for the interpretation of the MIR band polarized along the
conducting crystal direction. (2) In view of the anisotropic
compressibility, external pressure is a means to continuously
tune the electronic anisotropy in the $ab$ plane and to explore
the possibility of a crossover from one- to two-dimensional
behavior. A further question is whether the pronounced optical
anisotropy of the ambient-pressure phase is preserved across
the first-order structural phase transition near 18~GPa.

\begin{figure}[t]
\includegraphics[width=0.85\columnwidth]{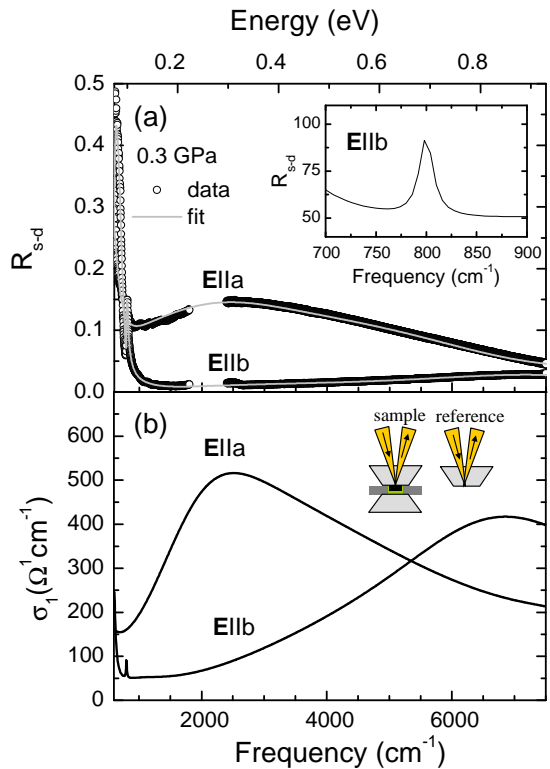}
\caption{(Color online) (a) Room-temperature reflectivity
spectra $R_{s-d}$ of LaTiO$_{3.41}$ inside the diamond anvil
cell at $P=0.3$~GPa for the polarizations
\textbf{E}$\parallel$\emph{a} and
\textbf{E}$\parallel$\emph{b}. The light grey lines are fits of
the reflectivity spectra with the Drude-Lorentz model, taking
into account the sample-diamond interface. Inset: Enlargement
of the low-frequency range of the \textbf{E}$\parallel$\emph{b}
reflectivity spectrum showing the optical phonon mode at 800
cm$^{-1}$. (b) Optical conductivity spectra obtained from the
Drude-Lorentz fits of the reflectivity data shown in (a).
Inset: Geometries for the sample and reference measurements.}
\label{Ref-fit}
\end{figure}

\begin{figure}[t]
\includegraphics[width=0.85\columnwidth]{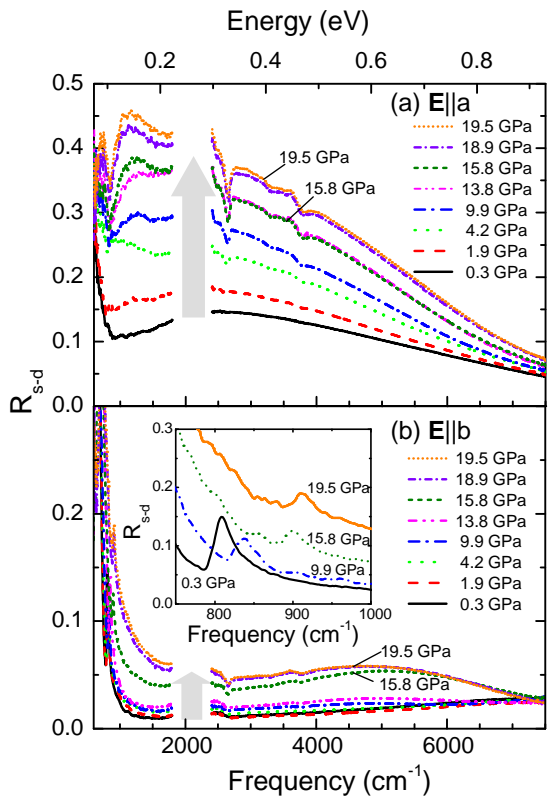}
\caption{(Color online) Room-temperature reflectivity spectra
R$_{s-d}$ of LaTiO$_{3.41}$ as a function of pressure for the
polari\-zation (a) \textbf{E}$\parallel$\emph{a} and (b)
\textbf{E}$\parallel$\emph{b}. The inset in (b) shows the
low-frequency range (700 - 1000~cm$^{-1}$) of the spectra for
four pressures. The phonon at around 800~cm$^{-1}$ shifts to
higher frequencies with increasing pressure. The arrows
indicate the changes with increasing pressure.} \label{R_s-d}
\end{figure}

\section{Experiment}
\label{sectionexperiment} The investigated LaTiO$_{3.41}$
crystals were grown by a floating zone melting process, and
their oxygen content was determined by thermogravimetric
analysis.\cite{Lichtenberg01} Pressure-dependent reflectance
measurements for the electrical field vector \textbf{E} of the
incident light along the \emph{a} and \emph{b} axes were
performed in the MIR frequency range (600-8000~cm$^{-1}$) at
room temperature, using a Bruker IFs 66v/S Fourier transform
infrared spectrometer. The measurements were carried out partly
at the University of Stuttgart and partly at the infrared
beamline of the synchrotron radiation source ANKA in Karlsruhe.
A diamond anvil cell equipped with type IIA diamonds suitable
for infrared measurements was used to generate pressures up to
20~GPa. To focus the infrared beam onto the small sample in the
pressure cell, a Bruker IR Scope II infrared microscope with a
15x magnification objective was used. A field stop of 0.6 mm
diameter was chosen, which yields a geometrical spot size of 40
$\mu$m on the sample (diffraction effects neglected). The
LaTiO$_{3.41}$ crystal was polished to a thickness of
$\approx$40 $\mu$m. The reflectivity of the free-standing
polished sample was checked and found to be in good agreement
with earlier results.\cite{Kuntscher03} A small piece of sample
(about 80 \nolinebreak $\mu$m x 80 \nolinebreak $\mu$m) was cut
and placed in the hole (150 $\mu$m diameter) of a steel gasket.
Finely ground KCl powder was added as a quasi-hydrostatic
pressure-transmitting medium. The ruby luminescence method
\cite{Mao86} was used for the pressure determination.

Polarized reflectivity spectra were measured at the interface
between sample and diamond. The measurement geometry is shown
in the inset of Fig.\ \ref{Ref-fit} (b). Spectra taken at the
inner diamond-air interface of the empty cell served as the
reference for normalization of the sample spectra. The absolute
reflectivity at the sample-diamond interface, denoted as
$R_{s-d}$, was calculated according to $R_{s-d}(\omega)=R_{\rm
dia}\times I_{s}(\omega)/I_{d}(\omega)$, where $I_s(\omega)$
denotes the intensity spectrum reflected from the
sample-diamond interface and $I_d(\omega)$ the reference
spectrum of the diamond-air interface. $R_{\rm dia}$ was
calculated from the refractive index of diamond $n_{\rm dia}$
to 0.167 and assumed to be independent of pressure. This is
justified because $n_{\rm dia}$ is known to change only very
little with pressure ($\Delta$$n_{\rm dia}$/$\Delta$P =
-0.00075/GPa).\cite{Eremets92,Ruoff94} Variations in
synchrotron source intensity were taken into account by
applying additional normalization procedures. Strain-induced
depolarization in the diamond anvil is considered negligible in
the pressure range covered in the present experiment.

\begin{figure}[t]
\includegraphics[width=0.75\columnwidth]{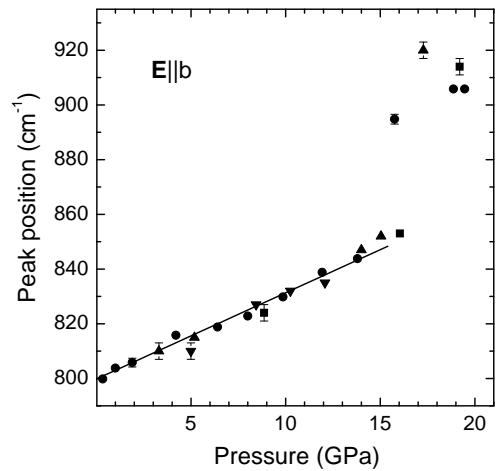}
\caption{Peak position of the phonon mode measured for
\textbf{E}$\parallel$\emph{b}. Different symbols are used for
different experimental runs. Up to 14~GPa the mode hardens in a
linear fashion, with a linear pressure coefficient of 3.2
cm$^{-1}$/GPa. The line is a linear fit of the data points. At
around 15~GPa an abrupt change in the frequency of the
vibrational feature occurs.} \label{Phonon_b}
\end{figure}

\begin{figure}[t]
\includegraphics[width=0.85\columnwidth]{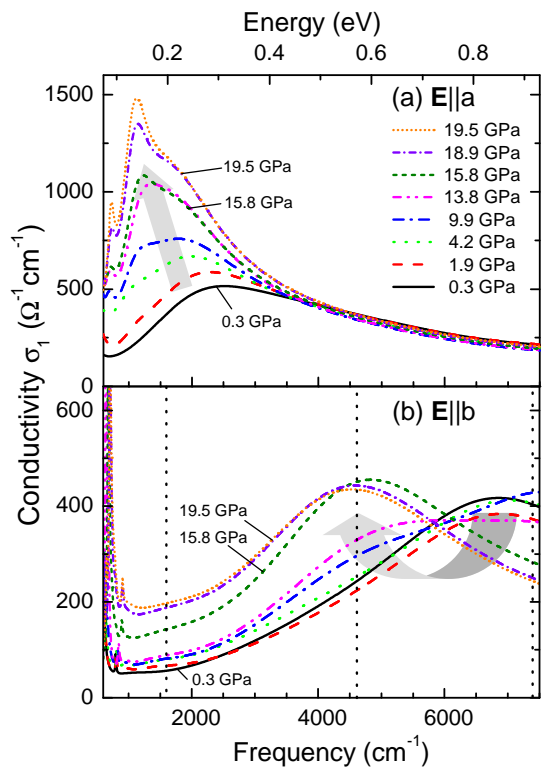}
\caption{(Color online) Pressure-dependent real part of the
optical conductivity of LaTiO$_{3.41}$ for (a) \textbf{E}$\parallel$\emph{a}
and (b) \textbf{E}$\parallel$\emph{b} at room temperature,
obtained by Drude-Lorentz fitting of the reflectivity data. The
arrows indicate the changes with increasing pressure. The dotted
vertical lines in (b) indicate the frequency positions analyzed in
Fig.\ \ref{changes}.}
\label{sigma1}
\end{figure}

\section{Results}
\label{sectionresults}

The reflectivity spectra of LaTiO$_{3.41}$ for the lowest
pressure (0.3~GPa) are shown in Fig.\ \ref{Ref-fit}(a) for
\textbf{E}$\parallel$\emph{a,b}. The region around
2000~cm$^{-1}$ is cut out from the experimental spectra since
the diamond multi-phonon absorption causes artifacts in this
range. The overall reflectivity of the sample in the diamond
anvil cell is lower than that of the free-standing sample
\cite{Kuntscher03} due to the smaller refractive index step at
the sample-diamond interface.

The optical conductivity was obtained by fitting the
reflectivity spectra with a Drude-Lorentz model combined with the
normal-incidence Fresnel equation
\begin{equation}
R_{s-d} =\left| \frac{n_{\rm dia}-\sqrt{\epsilon_s}}{n_{\rm
dia}+\sqrt{\epsilon_s}}\right|^2 , \epsilon_s = \epsilon_\infty +
\frac{i \sigma}{\epsilon_0 \omega} \quad ,
\end{equation}
where $\epsilon_s$ is the complex dielectric function of the
sample. With the 15x objective used in the
experiment, the angle of incidence of the radiation at the
diamond-sample interface ranges from 4.1$^{\circ}$ to 9.5$^{\circ}$; the
average deviation from normal incidence is thus small enough to
assume normal incidence for the data analysis. Furthermore, an
increase of the background dielectric constant (by 14.5\% at
maximum) according to the Clausius-Mossotti relation
\cite{Ashcroft76} was assumed in the Drude-Lorentz fits to account
for the pressure-induced reduction of the unit cell
volume.\cite{Loa04}

As an example, we present in Fig.\ \ref{Ref-fit}(a) the
Drude-Lorentz fits of the reflectivity spectra for the lowest
applied pressure (0.3~GPa); the resulting real part $\sigma_1$
of the optical conductivity is shown in Fig.\ \ref{Ref-fit}(b)
for both studied polarizations. For fitting the lowest-pressure
data, the fitting parameters for the reflectivity spectra of
the free-standing sample were used as starting parameters. The
resulting optical conductivity spectra at 0.3~GPa [Fig.\
\ref{Ref-fit}(b)] are in overall agreement with the
ambient-pressure results.\cite{Kuntscher03}

For the polarization of the radiation parallel to the
conducting axis $a$ the optical conductivity spectrum consists
of a pronounced, asymmetric absorption band, located at around
2500 cm$^{-1}$ at 0.3~GPa. For the perpendicular polarization
direction, {\bf E}$\parallel$$b$, a phonon mode located at 800
cm$^{-1}$ and a broad band centered around 7000 \nolinebreak
cm$^{-1}$ are observed for the lowest applied pressure.

The polarization-dependent reflectivity spectra R$_{s-d}$ for
pressures up to $\approx$20~GPa are presented in Fig.\
\ref{R_s-d}. Features around $\omega$=2500~cm$^{-1}$ and 3700
cm$^{-1}$ are artifacts originating from multi-phonon
absorptions of diamond which are not fully corrected by the
normalization procedure.
For \textbf{E}$\parallel$\emph{a} the reflectivity increases
continuously with increasing pressure in the whole frequency
range studied here. In contrast, for
\textbf{E}$\parallel$\emph{b} the overall reflectivity is
almost unchanged up to a pressure of $\approx$15~GPa and only
above 15~GPa R$_{s-d}$ increases strongly. Furthermore, the
phonon mode located at 800 cm$^{-1}$ in the
\textbf{E}$\parallel$\emph{b} spectrum hardens with increasing
pressure [for spectra see inset of Fig.\ \ref{R_s-d} (b)]. In
Fig.\ \ref{Phonon_b} the peak position of the phonon mode is
shown as a function of applied pressure: Up to 15 \nolinebreak
GPa the mode hardens in a linear fashion with a pressure
coefficient of $3.2\pm 0.5$~cm$^{-1}$/GPa; at around 15~GPa a
discontinuous change in the frequency of the observed spectral
feature occurs.

\begin{figure}[t]
\includegraphics[width=0.85\columnwidth]{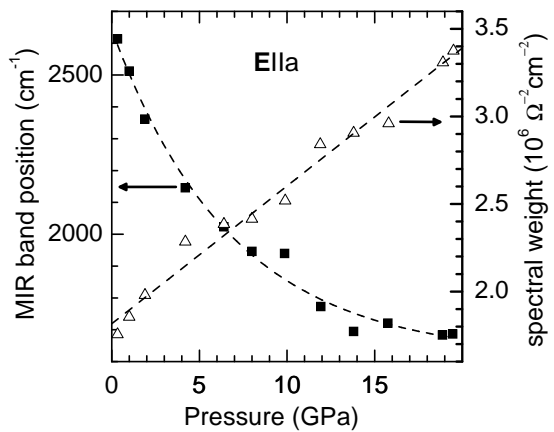}
\caption{Frequency position (filled squares) and spectral weight
(open triangles) of the {\bf E}$||$$a$ MIR band as a function of
pressure. Dashed lines are guides to the eye.}
\label{bandpos}
\end{figure}

The pressure-dependent real part $\sigma_1$ of the optical
conductivity obtained from the Drude-Lorentz fits is presented
in Fig.\ \ref{sigma1}. With increasing pressure the MIR band
observed in the {\bf E}$\parallel$$a$ optical conductivity
shifts to lower frequencies and its oscillator strength
increases. This MIR band is superimposed by a relatively narrow
peak at $\approx$1200~cm$^{-1}$, whose oscillator strength
strongly increases above 15~GPa. For the polarization {\bf
E}$\parallel$$b$ gradual changes set in at $\approx 10$~GPa.
With increasing pressure the absorption band located at around
7000~cm$^{-1}$ for the lowest pressure looses oscillator
strength and the spectral weight moves to the frequency range
4000 to 5000~cm$^{-1}$. A massive redistribution of spectral
weight occurs between 14 and 16~GPa. Furthermore, at around
$\approx15$~GPa a significant increase of the optical
conductivity in the low-frequency part ($<$2000 \nolinebreak
cm$^{-1}$) of the \textbf{E}$\parallel$\emph{b} spectrum is
observed.

Upon releasing pressure, the pressure-dependent trends, e.g.,
overall increase of reflectivity for {\bf E}$\parallel$$a$, the
sudden change in frequency of the {\bf E}$\parallel$$b$ phonon
mode and the redistribution of spectral weight in the high
frequency range, are reversible.

\section{Discussion} \label{sectiondiscussion}

\subsection{Low-pressure regime: $P<15$~GPa}

In the low-pressure regime ($P<15$~GPa) the changes in the
optical response with increasing pressure are conti\-nu\-ous.
In the {\bf E}$||$$b$ optical conductivity spectrum there is a
small redistribution of the high-frequency spectral weight near
$\approx$7000~cm$^{-1}$ towards lower frequency. That spectral
weight may be due to charge transfer excitations, and the
pressure-induced redistribution may reflect subtle alterations
in the crystal structure,\cite{Loa04} causing changes in the
electronic band structure.

For {\bf E}$||$$a$ one finds a monotonic redshift and spectral
weight growth of the pronounced MIR band with increasing
pressure. Based on Drude-Lorentz fits of the reflectivity
spectra, the contribution of the MIR band to the optical
conductivity was extracted. The zero crossing of the first
derivative of this contribution served as an estimate for the
frequency position of the band. The so-obtained position of the
{\bf E}$||$$a$ MIR band is plotted in Fig.\ \ref{bandpos} as a
function of applied pressure. In addition, we show the pressure
dependence of its spectral weight which increases by a factor
of two for pressures up to 20~GPa.

Based on its absolute strength and pressure dependence, the MIR band
can be interpreted in terms of (i) excitations of purely electronic
character and (ii) excitations involving electron-phonon coupling,
i.e., polaronic excitations.

For the interpretation of the MIR band in terms of purely
electronic excitations, it is instructive to compare the
ambient-pressure spectrum of LaTiO$_{3.41}$ with that of the
Mott-Hubbard insulator LaTiO$_{3.0}$. For LaTiO$_{3.0}$ the
increase of the optical conductivity at around 700 \nolinebreak
cm$^{-1}$ is due to excitations from the lower to the upper
Hubbard
band.\cite{Lunkenheimer03,Arima03,Katsufuji95,Okimoto95,Crandles94}
Upon hole doping, additional excitations within the
Mott-Hubbard gap were theoretically predicted
\cite{Jarrel95,Rozenberg96} and experimentally
demonstrated:\cite{Taguchi93,Okimoto95,Katsufuji95} namely, a
coherent Drude term and an incoherent MIR band due to
transitions between the quasi-particle peak at the Fermi energy
to the upper Hubbard band (or from the lower Hubbard band to
the quasi-particle peak).

In analogy, assuming a Mott-Hubbard picture, LaTiO$_{3.41}$
with an electronic configuration 3d$^{0.18}$ is in a highly
hole-doped regime, and the {\bf E}$||$$a$ MIR band could be
attributed to the predicted incoherent inner-gap excitations.
The observed pressure-induced redshift and spectral weight
growth of the MIR band reminds one of the doping- or
thermally-induced changes of the incoherent MIR band in
Mott-Hubbard systems. \cite{Imada98} Accordingly, the
pressure-induced effects in LaTiO$_{3.41}$ could be attributed
to the bandwidth-controlled delocalization of charges.

It is interesting to compare the crystal structure of
LaTiO$_{3.41}$ with that of the Mott-Hubbard system
LaTiO$_{3.0}$:\cite{Cwik03} At ambient conditions, the crystal
structure of LaTiO$_{3.0}$ is of the orthorhombic
GdFeO$_3$-type with characteristic tiltings and distortions of
the TiO$_6$ octahedra. The Ti-O1 bond length and Ti-O1-Ti bond
angle (with O1 denoting the apex oxygen ion) influence the $3d$
electron bandwidth,\cite{Taguchi93} and amount to 2.03 \AA\ and
154$^{\circ}$, respectively.\cite{Cwik03} A three-dimensional
network of tilted and distorted TiO$_6$ octahedra is also
present in LaTiO$_{3.41}$ within an ($a$,$b$)-slab. Each slab
consists of five chains of TiO$_6$ octahedra along the $a$ axis
and connected via their apical oxygen ions. Within a slab, the
octahedral tiltings and distortions are not homogeneous, but
the largest average Ti-O1-Ti bond angle (163$^{\circ}$) and
smallest average Ti-O1 bond length (1.99 \AA) are present
within the chain at the symmetrical position in the middle of
the slab. Interestingly, in LaTiO$_{3.0}$ a pressure-induced
insulator-to-metal transition is observed for Ti-O1 bond
lengths just below 2 \nolinebreak \AA. \cite{Loa05}
This suggests that in LaTiO$_{3.41}$ the central chains within
the slabs play a key role for the observed conducting properties
of this compound.

The other scenario for the interpretation of the pronounced
asymmetric {\bf E}$||$$a$ MIR absorption band of LaTiO$_{3.41}$
is in terms of an optical signature of polaronic
quasi-particles that are formed due to electron-phonon
interaction.\cite{Emin93} Within a polaronic picture, the
frequency position of the MIR band is a measure of the polaronic
binding energy and thus of the electron-phonon coupling
strength. Such a MIR absorption band was found in several
well-studied materials, like
cuprates,\cite{Bi93a,Falck93,Calvani96b,Lupi98,Lupi99}
manganites,\cite{Calvani96a,Kim98,Hartinger04} and
nickelates.\cite{Bi93a,Calvani96a,Bi93b,Crandles93} Also
several titanium oxides show the characteristic absorption
feature of polarons in the MIR frequency
range,\cite{Bogomolov68,Gerthsen65,Eagles84,Calvani93,Kabanov95}
and also for LaTiO$_{3.41}$ such a picture had been suggested
to explain the pronounced MIR band for {\bf
E}$||$$a$.\cite{Kuntscher03}

In the case of the manganites it was demonstrated that the MIR
band is very sensitive to the application of chemical or
external pressure, see e.g.
Refs.~\onlinecite{Loa01,Congeduti01,Postorino03}. In the case
of doped manganites\cite{Congeduti01,Postorino03}, pressure
effects were attributed to tuning the strength of the
electron-phonon coupling and thus the extent of the
localization of the charges. In general, a broadening of the
electronic bands, i.e., an enhancement of the electron
itineracy, and a stiffening of the lattice is expected upon
pressure application. As a consequence, the electron-phonon
coupling and therefore the polaronic binding energy should
decrease. So a shift of the MIR band to lower energies and an
increase of the oscillator strength, indicating an enhanced
delocalization of the charge carriers, are then expected. This
is in agreement with the observed pressure-induced changes of
the MIR absorption feature in LaTiO$_{3.41}$ [see Figs.\
\ref{sigma1}(a) and \ref{bandpos}].

\begin{figure}[t]
\includegraphics[width=0.85\columnwidth]{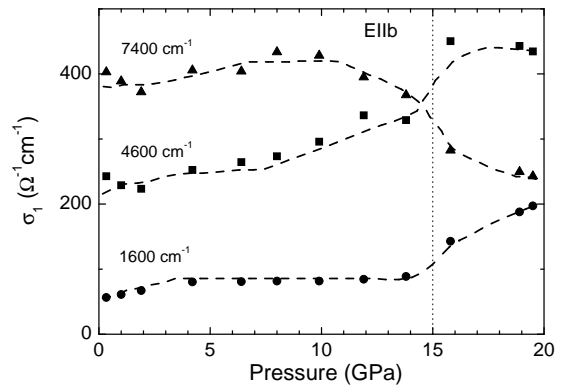}
\caption{Real part of the {\bf E}$||$$b$ optical conductivity
at three different frequencies (1600, 4600, and 7400~cm$^{-1}$)
as a function of pressure (extracted from Fig.\ \ref{sigma1}).
The dotted vertical line indicates the pressure where the
structural phase transition reported earlier \cite{Loa04}
occurs in the optical present optical study. Dashed lines are
guides to the eye.} \label{changes}
\end{figure}

Thus, based on the pressure dependence of the MIR band it is difficult to
draw a conclusion on the question whether this absorption feature 
is to be explained by a Mott-Hubbard or a polaronic scenario. 
For both the Mott-Hubbard and the polaron model several examples exist, 
where the distinct doping dependences of the spectral features were demonstrated.
\cite{{Katsufuji95,Taguchi93,Calvani96b,Lupi99,Bi93b}} Thus, also in the 
case of the titanate LaTiO$_{3.41}$ the doping dependence of the MIR 
absorption band might be an important additional piece of information.
Further insight could be obtained by a detailed lineshape analysis of 
the absorption band.

\subsection{High-pressure regime: $P>15$~GPa}

Near 15~GPa we observe discontinuous changes in the optical
response: (i) The oscillator strength of the narrow peak
superposing the MIR band for the po\-la\-ri\-zation {\bf
E}$||$$a$ increases. (ii) For {\bf E}$||$$b$ a pronounced
redistribution of spectral weight from high ($\approx$7400
cm$^{-1}$) to lower ($\approx$4600~cm$^{-1}$) frequencies
occurs, indicating the shift of the broad {\bf E}$||$$b$
excitation band. (iii) There is a sudden change in frequency of
the vibrational excitation showing up in the MIR spectra (see
Fig.\ \ref{Phonon_b}). (iv) In the low-frequency part
($\omega$$<$2000~cm$^{-1}$) the optical conductivity
perpendicular to the chains, {\bf E}$||$$b$, remains almost
constant in the lower pressure range, and starts to increase
above $\approx$15~GPa (see changes at 1600 cm$^{-1}$
illustrated in Fig.\ \ref{changes}).

These discontinuities are most probably related to the
pressure-induced structural phase transition, which was
observed at 18~GPa by x-ray diffraction measurements under more
hydrostatic conditions.\cite{Loa04} Due to the large number of
overlapping reflections, the high-pressure crystal structure
could not be determined; a reversible distortion of the
low-pressure crystal structure was suggested.\cite{Loa04}
According to our optical data, the structural transition
affects both the vibrational and electronic excitations of the
system. Above the phase transition, i.e., for $P>15$~GPa, the
broad {\bf E}$||$$b$ excitation band is located at
$\approx$5000~cm$^{-1}$ and remains almost unchanged when
increasing the pressure.

The abrupt increase of the low-frequency part
($\omega<2000$~cm$^{-1}$) of the optical conductivity spectrum
{\it perpendicular} to the chains deserves special attention.
It cannot be simply explained by the redshift of the
higher-lying broad band. An additional oscillator below
2000~cm$^{-1}$ needs to be included to describe the
high-pressure (i.e., $P>15$~GPa) reflectivity spectra with the
Drude-Lorentz model. This suggests the onset of a
pressure-induced dimensional crossover of the system at 15~GPa,
i.e., a significant increase of the hopping integral
perpendicular to the chains. However, the anisotropy of the
material is preserved up to the highest applied pressure
(19.5~GPa), since the overall optical conductivity for {\bf
E}$||$$a$ remains higher compared to {\bf E}$||$$b$.

\section{Summary}

We studied the polarization-dependent mid-infrared reflectivity
of the quasi-1D conducting titanate LaTiO$_{3.41}$ as a
function of pressure. Below 15~GPa the changes with increasing
pressure are continuous for both polarizations studied: For
{\bf E}$\parallel$$a$ the overall reflectivity increases; the
corresponding optical conductivity contains a pronounced MIR
absorption band showing a redshift and an increase of spectral
weight with increasing pressure. Based on its pressure
dependence, this MIR band can be interpreted in terms of
electronic transitions within a Mott-Hubbard picture in the
hole-doped regime, but the pressure-induced changes may also be
consistent with an interpretation in terms of polaronic
excitations. Additional information, like the doping dependence,
is called for, in order to be able to distinguish between the two 
possible scenarios.
For {\bf E}$\parallel$$b$ almost no change in the
reflectivity spectra is induced with increasing pressure up to
15~GPa; only a small redistribution of spectral weight towards
lower frequencies is observed.

Near 15~GPa, discontinuous changes are clearly observed in the
optical response. However, the optical anisotropy of the
low-pressure phase persists in the regime of the high-pressure
phase. This indicates a well-defined orientational relationship
between low-pressure and high-pressure phases. The dominant
change of the optical response at the phase transition occurs
for  {\bf E}$\parallel$$b$: a pronounced redshift of the
excitation band, a sudden change in frequency of the phonon
mode, and an increase of the low-frequency optical
conductivity. These changes can be related to the recently
observed\cite{Loa04} pressure-induced structural phase
transition, which alters the electronic band structure and
induces the onset of a dimensional crossover in this highly
anisotropic system.

\subsection*{Acknowledgements}
We thank M. Dressel, S. Schuppler, and H. Winter for fruitful
discussions and G. Untereiner for technical assistance. We
acknowledge the ANKA Angstr\"omquelle Karlsruhe for the provision
of beamtime and we would like to thank D. Moss, Y.-L. Mathis, 
B. Gasharova, and M. S\"upfle for assistance using the beamline ANKA-IR. 
Financial support by the  BMBF (project No.\ 13N6918/1) and the DFG (Emmy
Noether-program) is acknowledged.

\end{document}